\documentclass{article}

\usepackage{microtype}
\usepackage{graphicx}
\usepackage{subfigure}
\usepackage{booktabs} 

\usepackage{hyperref}

\usepackage[preprint]{icml2024}

\usepackage{amsmath}
\usepackage{amssymb}
\usepackage{mathtools}
\usepackage{amsthm}

\usepackage[capitalize,noabbrev]{cleveref}

\theoremstyle{plain}

\theoremstyle{definition}

\theoremstyle{remark}

\usepackage[textsize=tiny]{todonotes}

\icmltitlerunning{EVA-GAN: Enhanced Various Audio Generation via Scalable Generative Adversarial Networks}

\begin{document}

\twocolumn[
\icmltitle{EVA-GAN: Enhanced Various Audio Generation via \\ Scalable Generative Adversarial Networks}



\icmlsetsymbol{equal}{*}

\begin{icmlauthorlist}
\icmlauthor{Shijia Liao}{nvidia}
\icmlauthor{Shiyi Lan}{nvidia}
\icmlauthor{Arun George Zachariah}{nvidia}
\end{icmlauthorlist}

\icmlaffiliation{nvidia}{NVIDIA Cooperation, Santa Clara, United States}

\icmlcorrespondingauthor{Shijia Liao}{shijial@nvidia.com}
\icmlcorrespondingauthor{Shiyi Lan}{shiyil@nvidia.com}
\icmlcorrespondingauthor{Arun George Zachariah}{azachariah@nvidia.com}

\icmlkeywords{Machine Learning, ICML, Vocoder, TTS}

\vskip 0.3in
]



\printAffiliationsAndNotice{}  

\begin{abstract}
The advent of Large Models marks a new era in machine learning, significantly outperforming smaller models by leveraging vast datasets to capture and synthesize complex patterns. Despite these advancements, the exploration into scaling, especially in the audio generation domain, remains limited, with previous efforts didn't extend into the high-fidelity (HiFi) 44.1kHz domain and suffering from both spectral discontinuities and blurriness in the high-frequency domain, alongside a lack of robustness against out-of-domain data. These limitations restrict the applicability of models to diverse use cases, including music and singing generation. Our work introduces Enhanced Various Audio Generation via Scalable Generative Adversarial Networks (EVA-GAN), yields significant improvements over previous state-of-the-art in spectral and high-frequency reconstruction and robustness in out-of-domain data performance, enabling the generation of HiFi audios by employing an extensive dataset of 36,000 hours of 44.1kHz audio, a context-aware module, a Human-In-The-Loop artifact measurement toolkit, and expands the model to approximately 200 million parameters. Demonstrations of our work are available at {\selectfont\url{https://double-blind-eva-gan.cc}}.
\end{abstract}

\begin{figure*}[t!]
\vskip 0.2in
\begin{center}
\centerline{\includegraphics[width=0.7\textwidth]{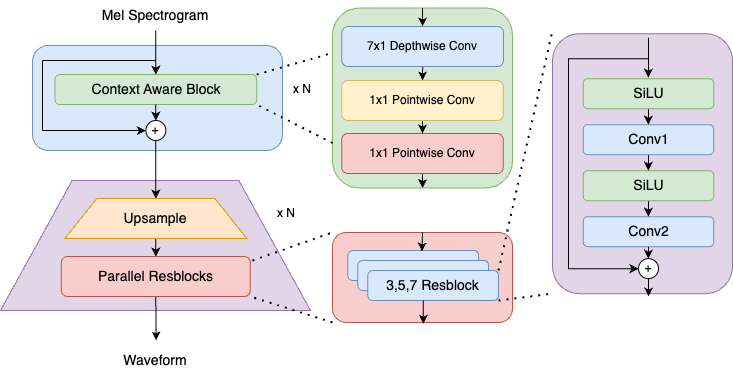}}
\caption{The EVA-GAN generator is composed of two main sections: Context Aware Blocks and Upsample Parallel Resblocks. The \textit{Context Aware Blocks}, a novel introduction in this paper, leverage residual connections and large convolution kernel to augment the context window and capacity of the module with minimal computational overhead. The \textit{Upsample Parallel Resblocks}, adapted from the HiFi-GAN's multi-receptive field fusion (MRF) blocks \cite{kong2020hifigan}, utilize the \textit{SiLU} \cite{elfwing2017sigmoidweighted} activation function for decoding features into a waveform.}
\label{fig:eva-gan-structure}
\end{center}
\vskip -0.2in
\end{figure*}

\section{Introduction}
\label{introduction}

Recently, GAN-based neural vocoders have revolutionized the generation of audio waveforms from acoustic properties, with broad applications in voice synthesis, voice conversion, and audio enhancement. Despite the efficiency in sampling and memory optimization offered by them, they face challenges such as \textbf{spectral discontinuities} and \textbf{blurriness} in the high-frequency domain, blocking the high-quality music and singing voice generation. Figure \ref{fig:compare} illustrates spectral disruptions when generating singing data using existing vocoders, including HiFiGAN \cite{kong2020hifigan} and BigVGAN \cite{lee2023bigvgan}, which represent the current state-of-the-art. We attribute these issues to a lack of data diversity, limited model capacity, a discrepancy between the context window sizes used during training and inference, and the absence of objective metrics for measuring these artifacts.

Another significant challenge in this domain is achieving high-quality audio fidelity, particularly used in music and singing synthesis, which remains insufficiently addressed. BigVGAN \cite{lee2023bigvgan} has proposed scaling of datasets and models as a pathway to state-of-the-art out-of-domain (OOD) performance. However, its reliance on the LibriTTS dataset \cite{zen2019libritts}, which is limited to low-fidelity (24kHz) speech data, falls short of capturing the diverse and rich acoustic properties required for realistic music and singing generation. While the scaling of models is recognized as a pivotal strategy for enhancing performance, the vast majority of vocoders are equipped with fewer than 20 million parameters. This limitation is primarily due to the significant computational expense and memory requirements associated with managing the gradient footprint, thus hampering advancements in the synthesis of high-fidelity music and singing audio.

Moreover, the prevalent evaluation of existing work on speech datasets, known for their reduced sensitivity to spectrogram quality, has led to a deficiency in effective objective metrics for the automatic assessment of high-frequency \textbf{artifacts} and spectrogram discontinuities in neural vocoders. We have identified that current metrics fail to detect artifacts that, despite being subtle (e.g., a discontinuity lasting only 20 milliseconds), are \textbf{highly perceptible} to humans. This underscores the necessity for a robust evaluation method capable of identifying such nuances during the training process.

The current issues in audio generation are very similar to the early challenges in Natural Language Processing (NLP). Prior to the advent of Large Language Models (LLMs) such as GPT-3, the consensus within the NLP field believed that solving its problems required specialized designs and techniques. Similarly, audio generation tasks such as singing synthesis, text-to-speech, and music synthesis have typically used different model architectures. However, the introduction of the GPT series marked a significant paradigm shift, illustrating that broad, scalable solutions could effectively tackle a range of NLP tasks without needing task-specific adjustments or unique model designs. Inspired by this shift, we present the Enhanced Various Audio Generation via Scalable Generative Adversarial Networks (EVA-GAN). By focusing on model and data scaling, EVA-GAN aims to modernize audio generation, creating a generalized and robust vocoder following industry-standard deep learning methods.

Our contributions are multifaceted:
\begin{enumerate}
    \item Compared to \cite{lee2023bigvgan}, we have expanded EVA-GAN to 200 million parameters and utilized a comprehensive 36,000-hour HiFi (44.1kHz) dataset, which are \textbf{largest} model and data used in Neural Vocoder to the best of our knowledge. 
    \item We introduce a novel context-aware module, referred to as CAM, which marks a significant leap forward in model performance, achieving outstanding advancements with virtually no additional computational burden, as documented in Table \ref{eva-gan-speed}.
    \item We propose a innovative training pipeline, which includes a longer context window (around 3 seconds), a loss balancer as initially introduced by Encodec \cite{défossez2022highencodec}, incorporated gradient checkpointing, and improved activation functions to boost training stability, reduce memory usage, and minimize the need for manual hyperparameter tuning.
    \item We build a brand new Human-In-The-Loop SMOS (Similarity Mean Option Score) evaluation toolkit, enables artifact monitoring and ensuring unparalleled alignment with human subjective perceptions.
\end{enumerate}

Overall, we created a state-of-the-art 44.1kHz Vocoder, EVA-GAN, especially suited for high-quality audio generation, establishing a new industry benchmark in this domain.

\section{Related Work}
\label{Related Works}

 Neural vocoders, which produce audio waveforms from acoustic properties using deep learning, have become indispensable in speech synthesis, voice conversion, and speech enhancement. The evolution of neural vocoder techniques can be segmented into three distinct phases: autoregressive (AR) or flow-based models \cite{van2016wavenet,kalchbrenner2018efficient,prenger2019waveglow}, GAN-centric approaches \cite{yamamoto2020parallel,yang2020multi,kong2020hifigan,jang2021univnet}, and direct spectral generation (DSG) strategies \cite{siuzdak2023vocos,kaneko2022istftnet}.

\begin{table*}[t!]
\caption{Speed measured across different models, results were obtained on a single A100 GPU, using a batch size of 16 with 512 frames (around 6 seconds) of audio in fp32 format. Comparing to original HiFi-GAN \cite{kong2020hifigan}, we have replaced leaky ReLU with SiLU \cite{elfwing2017sigmoidweighted}, applied in-place activation optimizations, and optimize kernel sizes for 44.1k generation. EVA-GAN-base is essentially HiFi-GAN-base enhanced with our Context Aware Module, while EVA-GAN-large represents a scaled version of the generator to 174M parameters. The reported training memory encompasses both forward and backward memory usage, excluding the application of the optimizer. Speed metrics are based on the average of 100 inferences, following 10 initial warm-up runs. PESQ (wide-band) are observed at LibriTTS dev set.}
\label{eva-gan-speed}
\vskip 0.15in
\begin{center}
\begin{small}
\begin{sc}
\begin{tabular}{lcccccc}
\toprule
Experiment & Total Params & Generator Params & Train Mem & Infer Mem & Infer Time & PESQ \\
\midrule
HiFi-GAN-base   & 13.6 M & 13.6 M & 43.2 GB & 2.1 GB & 177 ms & 3.5486 \\
EVA-GAN-base    & 34.85 M & 16.3 M & 46 GB & 2.2 GB & 193 ms & 4.0330 \\
EVA-GAN-big     & 192.99 M & 174.44 M & 68 GB & 2.7 GB & 402 ms & 4.3536 \\
\bottomrule
\end{tabular}
\end{sc}
\end{small}
\end{center}
\vskip -0.1in
\end{table*}

\subsection{GAN-based Neural Vocoders}
Of these, the current best-in-class GAN-based \cite{kong2020hifigan,lee2023bigvgan,kaneko2022istftnet,siuzdak2023vocos} vocoders boast impressive sampling efficiency and memory optimization. However, they come with their own set of challenges, like spectral inconsistencies, high and low frequency range artifacts, deconvolution checkerboard artifacts, leading to compromised audio quality. Furthermore, basic HiFi-GAN \cite{kong2020hifigan} tends to fall short when dealing with musical data, high-pitched audio, and out-of-distribution (OOD) content, causing audio disruptions.

\subsection{Development of Generator and Loss Terms}
UnivNet \cite{jang2021univnet} replaces the Mel-Spectrogram loss with a multi-scale STFT loss and adopts a multi-resolution discriminator to boost high-frequency domain recovery. Meanwhile, BigVGAN \cite{lee2023bigvgan} purposed to use Snake Activation to improve spectrogram quality and out-of-distribution (OOD) performance. On the other hand, models like NSF-HiFi-GAN \cite{Zhao2020nsf, kong2020hifigan, Openvpi_2022}, RefineGAN \cite{xu2021refinegan}, and SingGAN \cite{huang2022singgan}, incorporate an f0 source to elevate audio quality and spectral continuity, but this restricts them from utilizing large and varied datasets.

\subsection{Reduce Artifacts by Improving Discriminators}
Discriminators play a crucial role in vocoder training, striking a balance between reducing human-sensitive artifacts and optimizing objective scores, such as loss. Various approaches have been explored in previous works, including MPD, MSD, MRD, MS-STFTD, and MS-SBCQTD, all aiming to minimize these artifacts and enhance high-frequency domain reconstruction.

HiFi-GAN \cite{kong2020hifigan} introduced the Multi-Period Discriminator (MPD), transforming the waveform into 2D representations of varying heights and employing 2D convolutions to analyze periodic structures. In the same vein, HiFi-GAN's Multi-Scale Discriminator (MSD) processes the waveform into multiple 1D representations at different scales, enabling detailed analysis of time-domain information.

UnivNet \cite{jang2021univnet} proposed the Multi-Resolution Spectrogram Discriminator (MSRD or MRD), focusing on the multi-resolution time-frequency domain through the Short-Term Fourier Transform (STFT). Similarly, Encodec \cite{défossez2022highencodec} advocated for a Multi-Scale STFT Discriminator (MS-STFTD) to enhance audio generation quality.

Furthering this innovation, \cite{gu2023multiscalecqt} introduced the Multi-Scale Sub-Band Constant-Q Transform Discriminator (MS-SBCQTD). This novel approach supports the generator in more effectively restoring high-frequency components by utilizing the Constant Q Transform, an alternative to the conventional STFT.

\section{Preliminaries}
\label{Preliminaries}

This manuscript extends the foundational work on Generative Adversarial Network (GAN)-based vocoders, specifically leveraging the architectural paradigm introduced by HiFiGAN \cite{kong2020hifigan}. An exposition of the critical elements of this framework is imperative for understanding the subsequent developments. The inception of specific loss metrics by Least Squares GAN \cite{mao2017squares} heralded their adoption in GAN-based vocoders, prominently exemplified by HiFiGAN's implementation.

\subsection{Generator}
The generator within a GAN-based vocoder framework is tasked with the transformation of Mel-Spectrograms into unprocessed waveforms. The generator's loss metrics in HiFiGAN \cite{kong2020hifigan} encapsulate the Mel-Spectrogram loss $L_{mel}$, the adversarial loss $L_{adv}$, and the feature matching loss $L_{fm}$, articulated as follows:

\begin{equation}
    L_{mel}(G) = \mathop{\mathbb{E}} \Bigl[ \lVert \phi(x) - \phi(G(s)) \rVert_{1} \Bigr]
\end{equation}
\begin{equation}
    L_{adv}(G) = \mathop{\mathbb{E}} \Bigl[ (D(G(s)) - 1)^2 \Bigr]
\end{equation}
\begin{equation}
    L_{fm}(G) = \mathop{\mathbb{E}} \Bigl[ \sum_{i=1}^{T} \frac{1}{N_i} \lVert D^i(x) - D^i(G(s)) \rVert_{1} \Bigr]
\end{equation}
where $x$ represents the ground truth, $\phi$ symbolizes the Mel-Spectrogram function, $s$ denotes the Mel-Spectrogram corresponding to the audio signal, $T$ denotes the number of layers in the discriminator, $D$ and $G$ respectively signify the discriminator and generator, and $N_i$ denotes the number of features in the $i$-th layer of the discriminator

\subsection{Discriminator}

Beyond the Mel-Spectrogram loss, contemporary GAN-based vocoders incorporate multiple discriminators to attenuate perceptual distortions, which, despite potentially enhancing objective measures such as spectrogram fidelity, remain perceptibly discernible. Notably, these discriminators universally adopt multi-resolution feature analysis. 

Each discriminator is evaluated using the following adversarial loss metric:
\begin{equation}
    L_{adv}(D) = \mathop{\mathbb{E}} \Bigl[ (D(x) - 1)^2 + (D(G(s)))^2 \Bigr],
\end{equation}
where $x$ denotes the ground truth, $s$ represents the Mel-Spectrogram associated with the audio signal, and $D$ and $G$ respectively indicate the Discriminator and Generator.

\section{EVA-GAN}
\label{EVA-GAN}

We position EVA-GAN as an advancement over HiFi-GAN \cite{kong2020hifigan}, characterized by a larger context window, an improved structure, increased capacity, and an expanded dataset.

\subsection{Data Scaling}

\begin{table*}[t!]
\caption{Objective evaluation of EVA-GAN was performed on the LibriTTS development sets, and subjective evaluation was carried out using test sets akin to those in the BigVGAN study \cite{lee2023bigvgan}. To align with these benchmarks, EVA-GAN's outputs were down-sampled to 24kHz. The Vocos \cite{siuzdak2023vocos} weights were obtained from \href{https://github.com/gemelo-ai/vocos}{gemelo-ai/vocos}. UnivNet-c32 \cite{jang2021univnet} model weights were sourced from \href{https://github.com/maum-ai/univnet}{maum-ai/univnet}. The HiFi-GAN \cite{kong2020hifigan} (V1) version, trained on LJSpeech, VCTK, and LibriTTS at 22kHz, was sourced from \href{https://github.com/jik876/hifi-gan}{jik876/hifi-gan}. For BigVGAN, objective results were taken directly from the paper \cite{lee2023bigvgan}, and subjective evaluations were based on weights trained for 5 million steps, acquired from \href{https://github.com/NVIDIA/BigVGAN}{NVIDIA/BigVGAN}. }
\label{table:libritts}
\vskip 0.15in
\begin{center}
\begin{small}
\begin{sc}
\begin{tabular}{l|ccc|c|c}
\toprule
LibriTTS & M-STFT ($\downarrow$) & Periodicity ($\downarrow$) & V/UV F1 ($\uparrow$) & PESQ ($\uparrow$) & SMOS ($\uparrow$)  \\
\midrule
Ground Truth   & - & - & - & - & 4.909 \\
\midrule
Vocos   & 0.8580 & 0.1103 & 0.9555 & 3.6328 & 4.8577 \\
UnivNet-c32   & 0.8959 & 0.1333 & 0.9444 & 3.2566 & 4.8042 \\
HiFi-GAN (V1)   & 1.3647 & 0.1600 & 0.9309 & 2.9110 & 4.7596 \\
\midrule
BigVGAN-base   & 0.8788 & 0.1287 & 0.9459 & 3.5190 & 4.8545 \\
BigVGAN-big   & 0.7997 & 0.1018 & 0.9598 & 4.0270 & 4.8786 \\
\midrule
HiFi-GAN-base   & 1.0269 & 0.1230 & 0.9523 & 3.5485 & 4.8345 \\
EVA-GAN-base    & 0.9485 & 0.0942 & 0.9658 & 4.0330 & 4.8687 \\
EVA-GAN-big     & \textbf{0.7982} & \textbf{0.0751} & \textbf{0.9745} & \textbf{4.3536} & \textbf{4.9134} \\
\bottomrule
\end{tabular}
\end{sc}
\end{small}
\end{center}
\vskip -0.1in
\end{table*}

Traditionally, the training of a vocoder involves using datasets such as LJSpeech \cite{ljspeech17}, LibriTTS \cite{zen2019libritts}, VCTK \cite{Veaux2017CSTRVCvctk}, and M4Singer \cite{zhang2022msinger}. These datasets, however, either suffer from a low sampling rate (24k) or lack diversity, covering only a limited range of speakers and languages. The importance of data scaling was underscored by the BigVGAN study \cite{lee2023bigvgan}, which showed significant out-of-distribution improvements when scaled to the LibriTTS \cite{zen2019libritts} dataset. Building on this, we aim to further increase the scale to enhance model robustness.

To this end, the \href{https://github.com/fishaudio}{Fish Audio} community compiled a comprehensive 16,000-hour high-fidelity music and song dataset from \href{https://music.youtube.com/}{YouTube Music} and \href{https://music.163.com/}{Netease Music} (dubbed \href{https://huggingface.co/datasets/fish-audio-private/hifi-16kh}{HiFi-16000h}). This dataset encompasses a variety of languages including Chinese, English, and Japanese, and features a wide spectrum of musical instruments. To our knowledge, this is the largest high-fidelity audio dataset to date, effectively addressing out-of-domain sample concerns.

Furthermore, to boost speech performance, an additional 20,000 hours of diverse language audio from the broadcast platform \href{https://player.fm}{Player FM} (termed PlayerFM-20000h) was added by the \href{https://github.com/fishaudio}{Fish Audio} community. A balanced distribution of 50\% from HiFi-16000h and 50\% from PlayerFM-20000h was maintained to ensure sample diversity.

Our results highlight the critical role of scale and diversity in training datasets. Our baseline HiFi-GAN, trained on this extensive 36,000-hour dataset, effectively reduces spectral discontinuities and demonstrates the capability to replicate a wide array of audio types. This includes, but is not limited to, singing, speaking, bass, piano, and unique sounds like helicopter noise and boiling kettle sounds.

\subsection{Model Scaling}
Echoing BigVGAN's findings \cite{lee2023bigvgan}, which highlight the superior performance of larger models over smaller ones even on the relatively modest-sized LibriTTS dataset \cite{zen2019libritts} (about 1000 hours), we observed marked improvements in robustness and overall performance with an enlarged generator. Accordingly, we have scaled up the generator in our base EVA-GAN-base model from 16.3 million to 174.4 million parameters, thus creating the more potent EVA-GAN-big variant. The enhanced capabilities of EVA-GAN-big are evident in Table \ref{eva-gan-speed} and demonstrated in Figure \ref{fig:compare}, particularly in terms of continuousness and resolution in the high-frequency domain.

Scaling up discriminators, such as MPD \cite{kong2020hifigan} and MRD \cite{jang2021univnet}, did not yield proportionate benefits. Additionally, incorporating new discriminators like MS-STFT \cite{défossez2022highencodec} and MS-SBCQTD \cite{gu2023multiscalecqt} did not further enhance model performance. We hypothesize that this outcome is attributable to the considerable capacity of our model, which effectively captures subtle distinctions between the ground truth and generated audio across both low and high-frequency ranges. This capability negates the need for any trade-offs or overemphasis on a particular frequency range in the loss function.

\subsection{The Free Lunch: Context Aware Module (CAM)}
Despite the increased demand for computational resources and memory when scaling up the HiFi-GAN \cite{kong2020hifigan} generator, due to the challenges posed by long audio sequences and the substantial gradient footprint, we discovered that a 1D Convolution-based context-aware module namely CAM, utilizing the building blocks derived from ConvNeXt \cite{liu2022convnet}, requires significantly less memory and computational resources. This module not only is resource-efficient but also delivers notable improvements in both objective and subjective assessments, as demonstrated in Table \ref{eva-gan-speed}.

\subsection{Renew Training Paradigm}

While intuitively scaling the model, data, and context length can enhance performance, the challenge lies in achieving this within the bounds of limited computational resources and maintaining training stability. BigVGAN, previously the largest vocoder \cite{lee2023bigvgan}, encountered obstacles in further scaling due to training instability and computational constraints. To mitigate these issues, they reduced the context length, resulting in a compromise between stability and resource utilization — less resource-intensive, yet at the cost of stability.

Our investigation revealed a gap in the implementation of several efficient techniques in existing neural vocoders. 


\subsubsection{Larger Context Window}
In comparison to the context windows (typically 32 or 64 frames) used in HiFi-GAN \cite{kong2020hifigan} and BigVGAN \cite{lee2023bigvgan}, extending the window to 256 frames proved highly beneficial. This increase aids in faster model convergence, higher GPU utilization, significantly reduces spectrogram discontinuities. However, this expansion also raises GPU memory consumption and slows down the training speed. To mitigate this, we implemented several optimizations as described below.

\begin{figure*}[t!]
\vskip 0.2in
\begin{center}
\centerline{\includegraphics[width=0.95\textwidth]{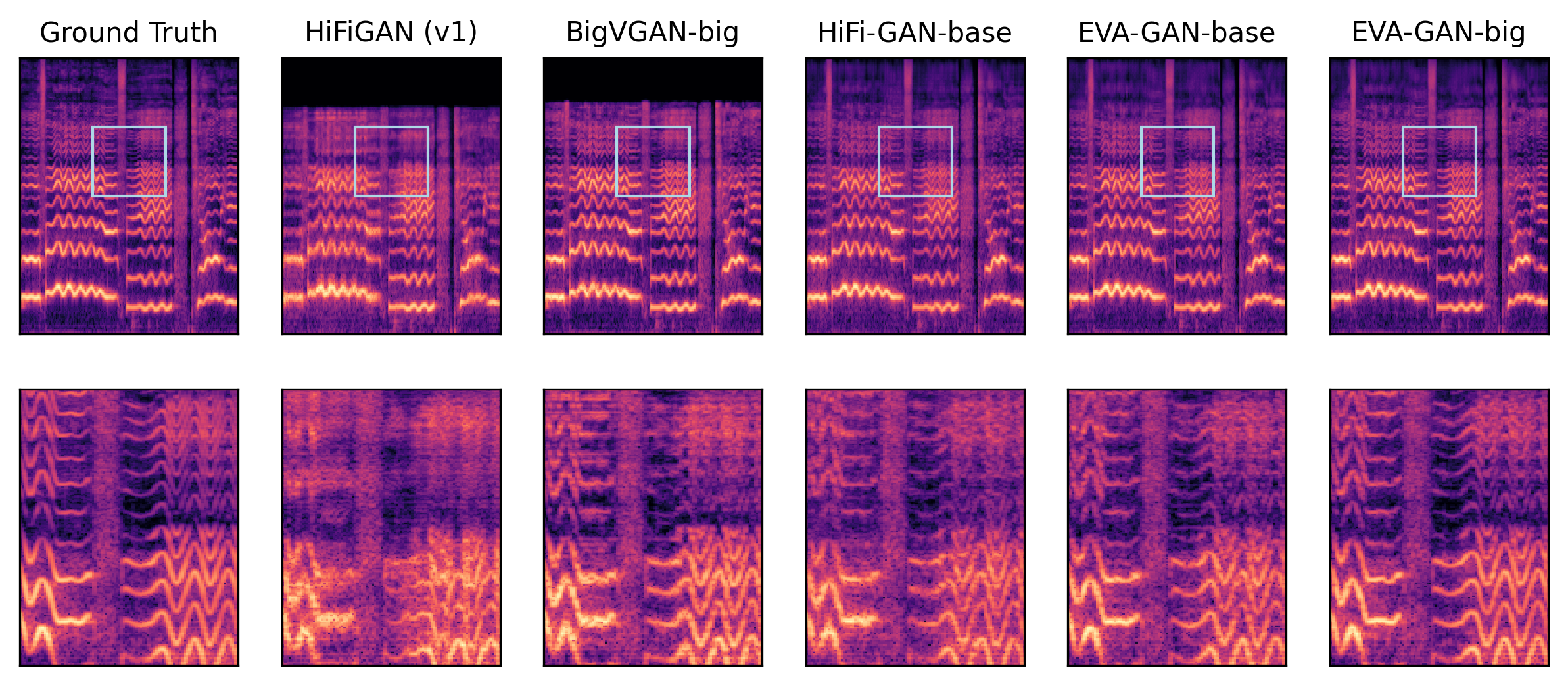}}
\caption{Spectrogram visualizations for a 44.1kHz singing voice generated by HiFi-GAN \cite{kong2020hifigan}, BigVGAN \cite{lee2023bigvgan}, and both the base and big versions of our EVA-GAN are presented, including zoomed-in views on high-frequency regions to illustrate differences in spectrogram continuity and high-frequency detail. The HiFi-GAN \cite{kong2020hifigan} (V1) model, trained on the LJSpeech, VCTK, and LibriTTS datasets at 22kHz, was obtained from \href{https://github.com/jik876/hifi-gan}{jik876/hifi-gan}. Weights for BigVGAN were sourced from the official repository \href{https://github.com/NVIDIA/BigVGAN}{NVIDIA/BigVGAN}.}
\label{fig:compare}
\end{center}
\vskip -0.2in
\end{figure*}

\subsubsection{SiLU in-place Activation}
We observed that replacing the Leaky ReLU activation function with SiLU \cite{elfwing2017sigmoidweighted} not only accelerates model convergence but also preserves final performance. Additionally, employing in-place activations for both the generator and discriminators wherever possible resulted in approximately a 30\% reduction in GPU memory usage.

\subsubsection{Gradient Checkpointing}
As indicated in Table \ref{eva-gan-speed}, large batch sizes are not feasible with a 256-frame context window, even on A100 GPUs. Thus, we applied gradient checkpointing \cite{chen2016traininggradientckpt} to both the generator and discriminators, significantly reducing the gradient memory footprint. For instance, the memory usage of the EVA-GAN-base generator decreased from 46GB to 16GB, albeit with a 30\% reduction in training speed.

\subsubsection{TensorFloat-32}
While mixed precision training with fp16 and bf16 is common in large language models and computer vision tasks, we encountered instability with fp16 (notably large gradient norms) and performance degradation with bf16 due to its lower precision. Therefore, utilizing A100 GPUs, we opted for TensorFloat-32 for training, which offers roughly twice the speed of fp32.

\subsubsection{Loss Balancer}
Upon scaling the model and adopting a 44.1k configuration, we encountered challenges in balancing various losses: feature matching loss, adversarial loss, Mel-Spectrogram loss, and multi-scale STFT loss. The loss balancer, a concept brought into the vocoder field by Encodec \cite{défossez2022highencodec}, emerged as a solution. This technique automatically balances these losses based on their gradients, allowing each to contribute equally to the parameter update process. In the absence of this balancing, we observed significant human-perceptible high-frequency artifacts, despite seemingly acceptable objective results such as l1 Mel-Spectrogram distance, attributable to the inability to properly adjust the weight of the discriminator loss.

\subsection{Human-In-The-Loop Artifact Measurement}

While metrics such as PESQ \cite{pesq} and Mel distance quantify the discrepancy between generated audios and ground truth, their high scores do not necessarily correlate with subjective evaluation outcomes. This misalignment arises because these metrics inadequately capture artifacts particularly perceptible to human listeners, especially those occurring in the high-frequency domain or as short-term spectrogram disruptions lasting only a few milliseconds. These nuances, though critical for audio quality, are overlooked by conventional metrics. In addressing this gap, discriminators in models like HiFiGAN \cite{kong2020hifigan} have been instrumental, highlighting their capability to discern such subtle differences, albeit not always reflected in objective metrics. 

To tackle these challenges, we have devised a Human-In-The-Loop Artifact Measurement toolkit, incorporating a SMOS (Similarity Mean Option Score) annotating tool and CLI, to continuously monitor and evaluate the quality of generated audio against human perceptual standards.

\section{Experiments Setup}

Adopting a similar approach to BigVGAN \cite{lee2023bigvgan}, we set our base learning rate at 1e-4 and applied gradient clipping at a norm of 1000 to manage gradient spiking and exploding. Our optimizer of choice was AdamW, with betas set at $[0.8, 0.99]$. In line with what is commonly practiced in LLMs, we opted for a step-wise learning rate decay, as opposed to the epoch-based approach used in HiFi-GAN \cite{kong2020hifigan} and its derivatives. Specifically, we implemented an exponential learning rate decay of $0.999999$ after each step.

Our model utilizes a 512x upsampling rate to accommodate a 160 bins Mel-Spectrogram input, with 2048 n\_fft, 512 hop length, and 2048 win length. As previously mentioned, we found that a larger context window is highly beneficial, so we used a 256 frames context window (approximately 3 seconds) instead of the 32 or 64 frames used in earlier models like HiFi-GAN \cite{kong2020hifigan} and BigVGAN \cite{lee2023bigvgan}. Due to memory constraints, our total batch size was set at 16, as opposed to 32 used in prior models. However, given the larger frame size of our model, we still achieved a higher effective frame count per batch, further enhancing training stability.

For our Multi Period Discriminator, as introduced by HiFiGAN \cite{kong2020hifigan}, we used periods of $[3, 5, 7, 11, 17, 23, 37]$. The Multi Resolution Discriminator and Multi Resolution STFT loss, as introduced by UnivNet \cite{jang2021univnet}, were set to resolutions of [[2048, 512, 2048], [1024, 120, 600], [2048, 240, 1200], [4096, 480, 2400], [512, 50, 240]]. This setup, comparing to the 24k configuration provided by previous works \cite{kong2020hifigan,lee2023bigvgan,jang2021univnet,siuzdak2023vocos}, improved the model's performance in the high-frequency domain.

With the aforementioned setup and using tf32 precision, we experienced stable training without any issues of gradient exploding or crashing. Unless specified otherwise, all models were trained on the 36,000-hour dataset comprising 50\% HiFi-16000h and 50\% PlayerFM-20000h for 1 million steps on NVIDIA A100 GPUs.

Starting with the HiFiGAN \cite{kong2020hifigan} v1 configuration, we replaced the Leaky ReLU activation with SiLU activation and established a 44.1k baseline, resulting in the \textbf{HiFi-GAN-Base}. We chose upsample rates of $[8, 8, 2, 2, 2]$, upsample kernel sizes of $[16, 16, 4, 4, 4]$, and parallel block (MRF blocks \cite{kong2020hifigan}) kernel sizes of $[3, 7, 11]$. This configuration led to a 13.6M generator.

Building upon \textbf{HiFi-GAN-Base}, we added the Context Aware Module before the generator: depths and dims for each block at $[3, 3, 9, 3]$ and $[128, 256, 384, 512]$, a 0.2 drop path ratio, and a kernel size of 7. This resulted in \textbf{EVA-GAN-base} with a 16.3M generator, considering the increased input dimension from 160 to 512, and an additional 18.6M for the Context Aware Module.

To develop \textbf{EVA-GAN-big}, we retained all settings from \textbf{EVA-GAN-base} but altered the upsample rate to $[4, 4, 2, 2, 2, 2, 2]$, the upsample kernel sizes to $[8, 8, 4, 4, 4, 4, 4]$, and the parallel block kernel sizes to $[3, 7, 11, 13]$. The initial channels for upsampling were set at 1536. This scaling increased the generator size to 174.4M, bringing the total parameter count to 193M.

\section{Results}
\label{Results}

\begin{table*}[t!]
\caption{SMOS evaluated on DSD100 test sets. BigVGAN \cite{lee2023bigvgan} weights, trained for 5 million steps, were acquired from the official repository \href{https://github.com/NVIDIA/BigVGAN}{NVIDIA/BigVGAN}. All samples are resampled to 44.1kHz for listeners.}
\label{table:dsd-100}
\vskip 0.15in
\begin{center}
\begin{small}
\begin{sc}
\begin{tabular}{l|ccccc}
\toprule
DSD100 & Mixture ($\uparrow$) & Bass ($\uparrow$) & Drums ($\uparrow$) & Vocals ($\uparrow$) & Other ($\uparrow$)  \\
\midrule
Ground Truth   & 4.7778 & 4.7237 & 4.8533 & 4.7531 & 4.8313 \\
\midrule
Vocos          & 3.1519 & 3.2716 & 3.9857 & 4.2078 & 3.0694 \\
UnivNet-c32    & 2.2778 & 2.9241 & 3.5507 & 3.2963 & 2.3247 \\
HiFi-GAN (V1)  & 2.4023 & 3.0833 & 3.5821 & 3.3188 & 2.5769 \\
\midrule
BigVGAN-base   & 3.3537 & 3.2821 & 4.2464 & 4.3846 & 3.6892 \\
BigVGAN-big    & 4.0854 & 3.8642 & 4.0909 & 4.5000 & 3.9747 \\
\midrule
HiFi-GAN-base  & 4.5658 & 4.1940 & \textbf{4.5493} & 4.6944 & 4.4605 \\
EVA-GAN-base   & 4.4133 & 4.2405 & 4.5467 & 4.6627 & 4.5634 \\
EVA-GAN-big    & \textbf{4.6197} & \textbf{4.4675} & 4.4658 & \textbf{4.7467} & \textbf{4.6053} \\
\bottomrule
\end{tabular}
\end{sc}
\end{small}
\end{center}
\vskip -0.1in
\end{table*}

We evaluated the performance of EVA-GAN and compared it with existing methods across multiple tasks, including LibriTTS (24k speech) and DSD-100 (48k music), and observed significant performance improvements.

Table \ref{eva-gan-speed} demonstrates that, compared to our optimized HiFi-GAN \cite{kong2020hifigan} baseline, the Context Aware Module adds minimal overhead in inference speed and training memory, yet significantly improves performance when trained on our dataset. This suggests that the additional parameters efficiently enhance network capacity without requiring excessive resources. Furthermore, while EVA-GAN-big is six times larger than EVA-GAN-base, it only doubles the training memory and slows inference time by a factor of one, maintaining a speed 250 times faster than real-time with a batch size of 16.

\subsection{Evaluation Metrics}
Following BigVGAN's methodology \cite{lee2023bigvgan}, we employed the following objective metrics for our LibriTTS evaluation:

\begin{itemize}
    \item Multi-resolution STFT (M-STFT), as provided in Parallel WaveGAN \cite{yamamoto2020parallel}, using the open-source implementation from \href{https://github.com/csteinmetz1/auraloss}{csteinmetz1/auraloss} \cite{steinmetz2020auraloss}.
    \item Periodicity error (based on CREPE) and voiced / unvoiced classification F1 score (V/UV F1), highlighting a common artifact in neural vocoders without a source module. We used CARGAN \cite{morrison2022chunkedcargan} code available at \href{https://github.com/descriptinc/cargan}{descriptinc/cargan}.
    \item Perceptual evaluation of speech quality (PESQ) \cite{pesq}, using the well-known automated voice quality assessment tool's wide-band version (16,000 Hz), available at \href{https://github.com/ludlows/PESQ}{ludlows/PESQ}.
\end{itemize}

In line with BigVGAN \cite{lee2023bigvgan}, we observed that the 5-scale mean opinion score (MOS) does not accurately reflect each model's performance, as most received high scores and minor data noise could alter the average scores significantly. Since our task involves copy-synthesis for vocoders, we aim for the model to generate outputs closely resembling the input Mel-Spectrogram. In cases of noisy inputs, we do not want the neural vocoder to bias the audio quality towards the training data, which might result in artificially high listener scores. 

Therefore, we adopted a 5-scale similarity mean opinion score (SMOS) approach, similar to \cite{lee2023bigvgan}. Specifically, participants were provided with both reference audio (Ground Truth) and generated audio to assess the quality of the generated audio. All participants were required to wear headphones and listen to the audios in a quiet environment for more accurate ranking.

\subsection{LibriTTS}

In Table \ref{table:libritts}, we present both objective and subjective results on LibriTTS \cite{zen2019libritts}, a 24kHz speech dataset. The objective evaluation metrics M-STFT, Periodicity, V/UV F1, and PESQ were calculated on the LibriTTS development set, encompassing both clean and 'other' categories with unseen speakers in the training of BigVGAN \cite{lee2023bigvgan} and HiFiGAN \cite{kong2020hifigan}. For subjective evaluation, we conducted SMOS tests on a randomly selected set of 100 files from the LibriTTS test set. 

Our findings, detailed in Table \ref{table:libritts}, reveal that EVA-GAN, despite being natively 44.1kHz and not trained on LibriTTS, significantly outperforms the current state-of-the-art, BigVGAN \cite{lee2023bigvgan}, in all objective and subjective metrics after downsampling to 24kHz. Notably, HiFi-GAN-base, with only changes in the training recipe, activation function, and the use of our large dataset, shows remarkable improvement over the HiFiGAN \cite{kong2020hifigan} baseline. This underscores the importance of our new training strategy and a more diverse dataset.

EVA-GAN-base further enhances the objective results of HiFi-GAN-base by incorporating a CAM. When compared to BigVGAN-base, EVA-GAN-base achieves better performance in most objective metrics while using less memory and offering faster inference speeds. Our largest model, EVA-GAN-big, validates the efficacy of scaling up, surpassing all existing models even though EVA-GANs were not trained on LibriTTS.

\subsection{DSD-100}
To manage resource constraints, we employed a strategy of random sampling, choosing 10 tracks from each of the five categories in the DSD-100 \cite{SiSEC16dsd100} dataset of 44.1kHz mixed audios for testing: Mixture, Bass, Drums, Vocals, and Others. For each track, we selected a 5-second clip from a random non-silent section for our evaluation, with the findings detailed in Table \ref{table:dsd-100}. 

Analyzing this dataset offered a unique perspective on the resilience and efficacy of neural vocoders in music generation, given its encompassment of critical musical components. The inference drawn is that a model's proficiency on the DSD-100 dataset likely translates to enhanced performance in high-fidelity music and speech synthesis.

Additional results, including those for extremely OOD tasks, are available in our demos at \href{https://double-blind-eva-gan.cc}{Here}.

\subsection{Ablation Studies}

\textbf{New Training Recipe and Larger Dataset:} Compared to the HiFiGAN v1 \cite{kong2020hifigan} baseline, our HiFi-GAN-base significantly improves objective metrics by optimizing the training recipe and dataset, as discussed in Section \ref{EVA-GAN} and evidenced in Table \ref{table:libritts} and Figure \ref{fig:compare}. 

\textbf{Context Aware Module:} The introduction of the Context Aware Module in our EVA-GAN-base significantly enhanced its objective metrics, adding minimal overhead. This is demonstrated in Tables \ref{eva-gan-speed}, \ref{table:libritts}, \ref{table:dsd-100}, and Figure \ref{fig:compare}. 

\textbf{Larger Model:} The scaling up of EVA-GAN to create EVA-GAN-big, detailed in Tables \ref{table:libritts}, \ref{table:dsd-100}, and Figure \ref{fig:compare}, shows the effectiveness of increasing model size. EVA-GAN-big, which is six times larger than EVA-GAN-base but without any changes in training hyperparameters or dataset, achieves superior performance in both speech and music domains, and exhibits robustness across various types of audio. Additional results demonstrating its OOD performance can be found on our \href{https://double-blind-eva-gan.cc}{Here}.

\section{Conclusions}
\label{Conclusions}
In this paper, we introduced \textbf{EVA-GAN}, a groundbreaking audio generation model that sets new benchmarks in the realm of neural vocoders. By leveraging an extensive dataset and incorporating innovative features such as a CAM and scaling the model to around 200M parameters, \textbf{EVA-GAN} significantly outperforms existing models in terms of spectral continuity, high-frequency reconstruction, and robustness in out-of-distribution data performance.

Our comprehensive experiments, conducted on a diverse range of audio data, including the largest dataset to date, demonstrate \textbf{EVA-GAN}'s superior ability to generate high-quality, realistic audio across various domains. This achievement not only marks a significant advancement in audio synthesis but also opens new avenues for future research and applications in speech synthesis, music generation, and beyond.

\textbf{EVA-GAN}'s remarkable performance, underscored by its state-of-the-art results and efficiency, establishes it as the new gold standard in audio generation, promising to enhance a wide array of applications in the audio domain.

\bibliography{references}
\bibliographystyle{icml2024}




\end{document}